\documentclass{ringb99}
\usepackage{graphics}

\begin{document}
\title{Particle reacceleration in the Coma cluster and 
the radio, EUV and hard X--ray emissions.}
\author{G. Brunetti\inst{1,2} 
\and L. Feretti\inst{2} \and
 G. Giovannini\inst{3,2} \and G. Setti\inst{1,2}}  
\institute{Dip. di Astronomia, Univ. di Bologna, via Ranzani 1,
I--40127 Bologna, Italy
\and  
Istituto di Radioastronomia del CNR, via Gobetti 101,
I--40129 Bologna, Italy
\and
Dip. di Fisica, Univ. di Bologna, via Berti--Pichat 6/2, 
I--40127 Bologna, Italy}

\authorrunning{Brunetti et al.}

\titlerunning{Reacceleration in the Coma halo}

\maketitle

\begin{abstract}

The radio spectral index map of the Coma halo 
shows a progressive steepening of the spectral index
with increasing radius.
Such a steepening cannot be simply justified by
models involving continuous injection of fresh particles
in the Coma halo or by models involving diffusion of 
fresh electrons from the central regions.

We propose a scenario in which the relativistic 
particles injected in the Coma cluster by some 
processes (merger, accretion shocks, turbulence)
are systematically reaccelerated for a relatively long time
($\sim$ 1Gyr).
We show that such a scenario can account 
for the synchrotron radio spectral index distribution, 
for the radio halo size at different
frequencies,  and for the total (integrated) radio
spectrum of the Coma halo.
We also show that for a suitable choice of the 
parameters the model can also account for the
hard X--ray flux observed by BeppoSAX via the 
inverse Compton scattering of the cosmic microwave 
background radiation by the synchrotron electrons.
The possibility to account for the
EUV flux detected with the EUVE satellite by
the inverse Compton emission is also discussed.

\end{abstract}

\section{Introduction}

The Coma halo is the most famous example of a diffuse
radio emission in clusters of galaxies.

At the lowest frequencies the radio emission 
extends up to $\sim 30-40$ arcmin 
from the center (e.g. Hanisch \& Erickson 1980 at 43 MHz;
Henning 1989 at 30.9 MHz), while 
at higher frequencies detailed radio images have been
obtained at 327 MHz (Venturi et al. 1990) and 
at 1.38 GHz (Kim et al. 1990).
By applying Gaussian fits it has been found 
that the 327 MHz FWHM  
(28$\times$20 arcmin) is significantly
larger than that inferred at 1.38 GHz 
(18.7$\times$13.7 arcmin), but
smaller than the low frequency size.

The 327--1400 MHz spectral index map of the Coma halo
shows that the spectrum steepens rapidly with
increasing distance from the center (Giovannini et al.1993).

The Coma halo has also been observed with
the Effelsberg single--dish 100 m telescope
(Deiss et al. 1997, at 1.4 GHz; Schlickeiser et al. 1987,
at 2.7 GHz).
The 1.4 GHz observations reveal
 a flux density 
at large scales higher than that measured with
the synthesis aperture instruments and confirm a 
spectral index steepening with increasing radius. 
The 2.7 GHz measurements appear to be inconsistent with 
these results suggesting that 
the Schlickeiser et al. integrated
flux may be underestimated (Deiss et al. 1997). 

Lieu et al.(1996) have detected extreme ultraviolet emission  
(EUV) in excess to that expected by 
extrapolating downward the thermal X--ray spectrum
emitted by the intracluster gas (kT = 8.21 keV).
Recently, Fusco--Femiano et al.(1999)
have discovered an hard X--ray tail exceeding the
thermal emission.

Synchrotron emission in the radio band, 
inverse Compton (IC) emission in the EUV and hard X--rays 
from cluster of galaxies are expected in the
framework of continuous injection of primary  
relativistic electrons (Sarazin 1999; V\"{o}lk 
\& Atoyan 1999).
The models invoking a secondary production of the
relativistic electrons also predict 
 a large gamma--ray flux 
from neutral pion decay;
in the case of Coma, an IC origin of the hard 
X-ray tail would lead to a gamma--ray flux considerably  
larger than the EGRET upper limit 
(Blasi \& Colafrancesco 1999).

\noindent
In this paper we will assume $H_0 = 75$ km s$^{-1}$ Mpc$^{-1}$.

\section{A model with systematic reacceleration}

\subsection{The spectral steepening problem}

A model should be able to reproduce
the observed steepening in the radio spectral index distribution, 
the integrated radio spectrum and 
the brightness distribution
observed at different frequencies.

The spectral 
steepening with radius is probably the most difficult
to reproduce. 
It cannot be related to the diffusion of
fast ageing particles from the central part into  
the cluster volume, because
the diffusion velocity is relatively low 
(Berezinsky et al. 1997) 
and the diffusion time 
is too large compared to the radiative time of 
the relativistic electrons (Sarazin 1999).

Alternatively one might consider a continuous injection
of fresh particles in the cluster volume.
To justify the steep spectrum observed in the
peripheral regions of the Coma cluster one would 
have to assume a space
modulation of the injection rate such that the 
injection is progressively stopped 
moving from the periphery toward the center
in a crossing time smaller than the radiative time
of the emitting particles.
This appears to be difficult to achieve in the Coma cluster,
since the  physical mechanism responsible for
such a modulation should then propagate at a velocity
20--50 times larger than the sound speed.

It may be thought that these problems can be solved 
by assuming that the 
spectrum of the continuously injected particles
is steeper with increasing distance from the
cluster center.
However, it can be shown that
the superposition of the corresponding synchrotron spectra,
constrained by the
observed brightness distribution, 
leads to an integral spectrum flat at high frequencies
($\sim 0.6-0.7$) and steep at low frequencies ($\sim 1.4-1.7$),
in contrast with the
observations that do not show a substantial 
flattening of the spectrum at high frequencies.

\subsection{The effect of systematic reacceleration on the
particle spectrum}

In order to avoid the above mentioned difficulties,
we investigate the possibility that systematic
reacceleration is present in the cluster gas and 
propose a two phases scenario for the Coma halo.

During the first phase the particles are  
injected and reaccelerated over the cluster volume
(starting at $t_2$) 
for a relatively long time ($t_1-t_2  \sim$ 1--3 Gyrs) by a
major merger event; evidence of a recent merging has been
pointed out (see Briel et al. 1992 for a discussion on  
recent mergings). 
 We further assume that following the injection phase
the particles are being
systematically reaccelerated for a relatively short
time ($\tau \equiv t_0-t_1$) till the present cosmic time $t_0$  
by secondary accretion shocks and/or turbulence.

By assuming a power law injection spectrum 
($\propto q \gamma^{-delta}$),   
the evolution of the particle density during the injection 
phase can be obtained by solving the kinetic equation (Kardashev 1962):

\begin{equation}
{{\partial N(\gamma,t)}\over{\partial t}}=
{{\partial}\over{\partial \gamma}} 
[-2\beta \gamma+\alpha]
+ q\gamma^{-\delta}
\end{equation}

with $\beta$ the radiative losses 
($d\gamma/dt=-\beta \gamma^2$)
and $\alpha \sim 1/T_{a}$ the rate of 
systematic reacceleration. 

We assume that the main cooling mechanism for the
relativistic particles in the Coma cluster is the 
IC scattering with the CMB photons, implying an 
average strength of the magnetic field $B \leq 3 \mu$G
across the cluster.

The solution for the first phase, at time $t_1-t_2$, 
is given by:

\begin{eqnarray}
N(\gamma, t_2, t_1)= {q\over{\alpha(\delta-1)}}
\gamma^{-\delta} 
{{ 1-e^{-\alpha(t_1-t_2)} }\over{ 
[1 - {{\gamma}\over{\gamma^0_b}}
-e^{-\alpha (t_1-t_2)} ] }} \cdot \nonumber\\
\{ e^{(\delta-1)\alpha (t_1-t_2)} 
[1-{{\gamma}\over{\gamma^0_b}}]^{\delta-1} -1 \}
\end{eqnarray}

\noindent
for $\gamma\leq \gamma^0_b \equiv \alpha/ \{\beta 
(1-e^{-\alpha (t_1-t_2)}) \}$, and

\begin{equation}
N(\gamma, t_2, t_1)= {{ q \gamma^{-\delta} 
(1-e^{-\alpha(t_1-t_2)}) }\over{\alpha(\delta-1)}}
[{{\gamma}\over{\gamma^0_b}}
+e^{-\alpha(t_1-t_2)} -1]^{-1} 
\end{equation}

\noindent
for $\gamma > \gamma^0_b$.

If the particles are reaccelerated during the second phase
for a time $\tau=t_0-t_1$, from Eq.(2) 
(under our assumptions the majority of the particles 
follow Eq.2) the spectrum becomes:

\begin{eqnarray}
N(\gamma, \tau, t_2, t_1)=
{{q e^{\alpha \tau (\delta-1)} }
\over{\alpha (\delta-1)}}
(1-e^{-\alpha (t_1-t_2)}) \gamma^{-\delta}
(1- {{\gamma}\over{\gamma_b}})^{\delta-2} \nonumber\\
\cdot
\{1-e^{-\alpha (t_1-t_2)} - {{ \gamma }\over{ \gamma^0_b}}
e^{-\alpha \tau}
(1- {{\gamma}\over{\gamma_b}})^{-1}  \}^{-1} \cdot
\nonumber\\
\{ e^{(\delta-1)\alpha (t_1-t_2)} 
[1- {{ \gamma e^{-\alpha \tau}}\over
{(1- {{\gamma}\over{\gamma_b}} ) \gamma^0_b }} ]^{\delta-1}
-1 \}
\end{eqnarray} 

\noindent
for $\gamma\leq \gamma_b \equiv \alpha/ \{\beta 
(1-e^{-\alpha \tau}) \}$, being zero at higher energies.
  
If the reacceleration is
sufficiently efficient,  
i.e.$\tau >> 1/\alpha$, 
Eq.(4) approaches to:
\footnote{It should be noticed that Eq.(5) holds 
for an initial power law energy spectrum 
$f(\gamma) \propto p^{-\delta}$ without low energy
cut--off.
In the case of a low energy cut--off and/or Coulomb
losses the evolution of the particle energy distribution 
is much more complicated (Brunetti et al. 1999),
but the main results of the model are still  valid.}

\begin{equation}
N(\gamma, \tau) \rightarrow 
K \gamma^{-\delta}
(1- {{\gamma}\over{\gamma_b}} )^{\delta-2}
\end{equation}

\noindent
We remark that, if the second phase is
sufficiently long, the
reacceleration during the first 
phase is not necessary to obtain the
final shape.

Under the assumption that the electron momenta  are 
isotropically distributed and Eq.(5), 
the synchrotron emissivity  
is given by (Jaffe \& Perola,1973):

\begin{eqnarray}
j({{\nu}\over{\nu_b}})=
\sqrt{3} {{e^3}\over{mc^2}} K B \gamma_b^{1-\delta}
\int_0^{\pi/2} d\theta sin^2 \theta 
\cdot \nonumber\\
\int_0^1 dx
F\left({{\nu/\nu_b}\over{x^2 sin\theta}}\right) 
x^{-\delta} (1- x)^{\delta-2}
\end{eqnarray}

where $\theta$ is the pitch angle, and 
$\nu_b$ is the critical 
frequency ($\theta=90^o$, $\gamma=\gamma_b$).

\subsection{Modeling the Coma halo}

Deiss et al.(1997) derived relationships between the observed
spectral index and the intrinsic spectral index
distribution of the
synchrotron emission assuming a spherical symmetry of the
relevant physical quantities.
\begin{figure}
\resizebox{\hsize}{!}{\includegraphics{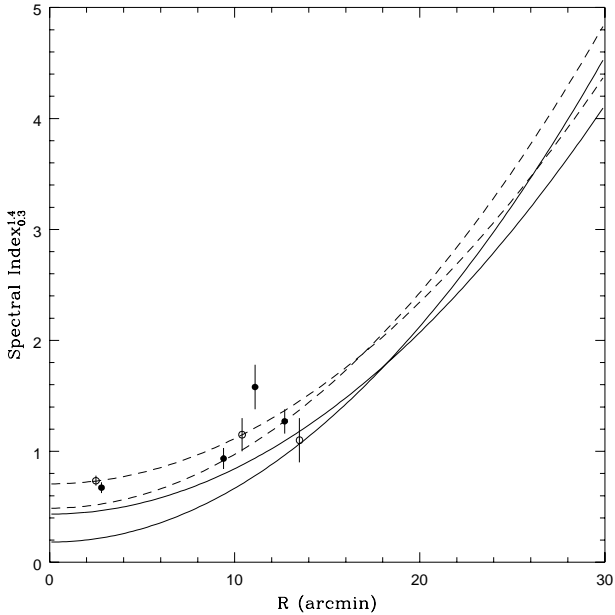}}
\caption[]{The predicted observable (dashed lines)
and intrinsic (solid lines) spectral index 
for the Coma halo
are reported as a function of the distance from the center.
The curves are obtained from Deiss et al.(1997) relationships
based on 327/1380 MHz (bottom dashed/solid lines
on left corner)
and on 327 (single dish)/ 1400 MHz observations.
The points are taken from two slices (open and filled 
circles respectively)
through the spectral index map (327--1380 MHz) 
of Giovannini et al.(1993).}
\end{figure}
By assuming a Gaussian spatial profile
of the emission coefficient they derive the 
intrinsic spectral index at a given radius and 
the observed spectral index at the same radius
(projected on the plane of the sky)
as a function of observable quantities.
In the case of the Coma halo they have shown
that the FWHMs and flux densities of the 
radio observations (Venturi et al. 1990; Kim et al.
1990) can be reproduced by a central intrinsic spectral index
$\sim 0.4$,
leading to a central observed spectral index of $\sim 0.6-0.8$,
in agreement with Giovannini et al. (1993) findings.
In Fig.1 we plot the 
observed and intrinsic spectral index predicted for the 
Coma halo from Deiss et al. (1997) relationships based on
the measurements at 327 MHz (Venturi et al.1990), 1.38 GHz
(Kim et al. 1990), and 1.4 GHz (Deiss et al. 1997).
The points in Fig. 1 are taken from two
different slices trough the Coma halo
(Giovannini et al. 1993).

From Eq.(6) we obtain the shape of the synchrotron spectrum 
as a function of the frequency measured in terms of the break
frequency $\nu_b$. 
By calculating the spectral index between two nominal
frequencies  
of constant ratio 1.4 GHz/327MHz, the break frequency 
$\nu_b(R) \propto B(R)\gamma_b^2(R)$ is obtained 
when the calculated matches the 
intrinsic spectral index of Fig.1.
The break frequency as a function of $R$ is represented in Fig.2  
for several values of the injected electron spectral 
index $\delta$.
\begin{figure}
 \resizebox{\hsize}{!}{\includegraphics{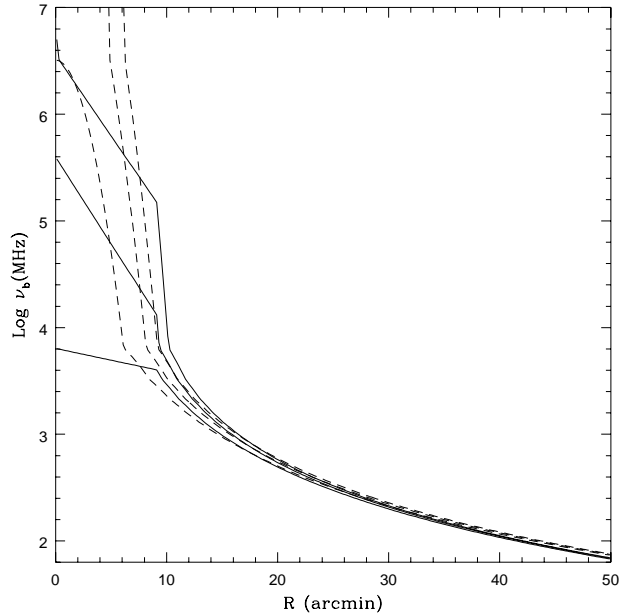}}
\caption[]{The calculated intrinsic break frequency is given
as a function of the distance $R$ from the cluster center.
The calculations are performed by assuming 
the 327--1380 MHz 
intrinsic spectral indices (dashed lines) as shown in Fig.1,
or the mean spectral index between 
327/1380 MHz and 327/(single dish) 1400 MHz 
measurements (solid lines; here a correction 
for non--spherical symmetry has been applied
to the Deiss et al.(1997) relationships).
The values are reported for three different $\delta$:
1.84, 2.01, 2.12 respectively from the bottom of
the diagram.}
\end{figure}
Due to the very flat intrinsic spectral index at 
the center of the 
Coma halo ($\alpha_{0.3}^{1.4} \sim 0.4$),  
the models with $\delta \geq 1.8-1.9$ cannot match the data
within $\sim 10$ arcmin from the center and
the break frequency goes to infinity (Fig.2).
A correction to Deiss et al. (1997) relationships, due to the
non perfect spherical symmetry of the Coma halo
(we assume it elongated on the plane of the sky),  
produces a steeper intrinsic radio spectral indices 
in the central part of the cluster allowing also 
the models with $\delta \sim 2$ to converge.
However, since we do not know the three--dimensional
structure of the Coma halo, the modeling of the
central $\sim$10 arcmin 
and the determination of the physical properties
remain uncertain.

As a zero order approximation
we assume that the large scale magnetic field 
$B$ is approximately constant over the volume of the cluster.
In this case, different break frequencies in the synchrotron
spectrum directly correspond  
to different break energies in the electron energy distribution,  
i.e. different reacceleration times 
($\alpha^{-1}$, $\gamma_b \rightarrow
\alpha/\beta$).
In Fig.3 we give the reacceleration time as a function
of $R$ for a number of
choices of the intracluster magnetic field.
In order to explain the very flat intrinsic radio spectrum
in the cluster center 
the efficiency of the reacceleration should rapidly
increase and the reacceleration time should become smaller
toward the central region,
whereas in the peripheral regions the reacceleration 
times are larger than $\sim 3 \cdot 10^7$yrs.
\begin{figure}
 \resizebox{\hsize}{!}{\includegraphics{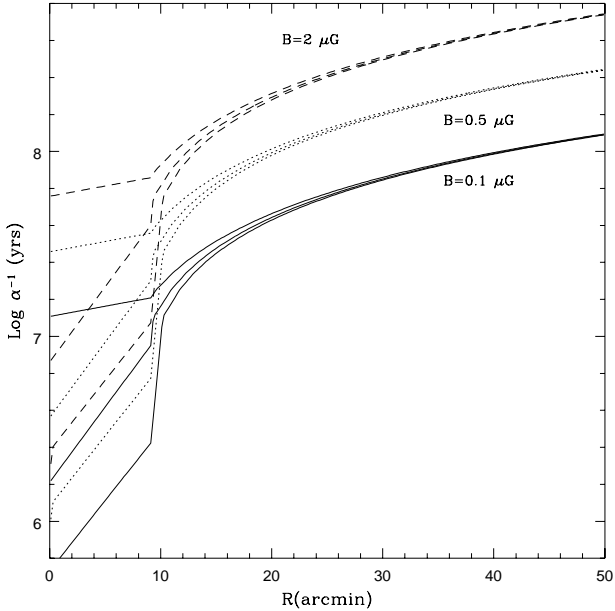}}
\caption[]{The reacceleration time is given as a function
of the distance $R$ from the center.
The calculations are reported for different magnetic field
strengths: 0.1 $\mu G$ (solid lines), 0.5 $\mu G$ (dotted lines),
and 2 $\mu G$ (dashed lines).
From each value of $B$, 
we plot three curves corresponding to
$\delta$=1.84, 2.01, and 2.12
starting from the bottom of the diagram.
As in Fig.2 a correction for
the non spherical symmetry of the Coma halo has been applied.}
\end{figure}
By integrating Eq.(6) along the line of sight we obtain 
the radio brightness at a given frequency
as a function of the projected distance from the center.
Given the magnetic field strength $B$ and the
geometry, the brightness is a function of the normalization
$K(R)$ of the emitting electron population.
The model should reproduce the FWHMs of the
radio brightness observed at different frequencies
(327 MHz, 1380 MHz, and 1400 MHz); by comparing the
predicted and observed FWHMs one obtains $K(R)$.
We find that a good representation is a $\beta$--model--type
function : $K(R) = K_0 \{1+ (R(arcmin)/17.5)^2 \}^{-1.5}$, where
$K_0$ depends on the assumed value of $B$.

The model at this point can reproduce both 
the observed and intrinsic synchrotron spectral
indices between 327 and 1400 MHz and  
the observed FWHMs of the brightness distribution at
these frequencies.
\begin{figure}
 \resizebox{\hsize}{!}{\includegraphics{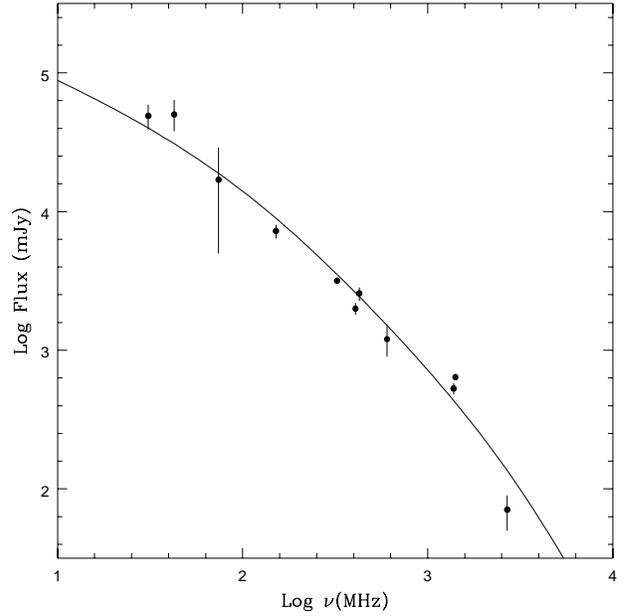}}
\caption[]{The calculated total radio spectrum of the
Coma halo is compared with the observed fluxes.
The calculation is performed by assuming $\delta=2.12$
(see the text).
The data are taken from Table 1 of Deiss et al.(1997).}
\end{figure}
A fundamental test is to check whether the model can
reproduce the shape of the total synchrotron spectrum
observed from the Coma halo.
We have then integrated Eq.(6),  
with the derived $\nu_b(R)$ (Fig.2, solid lines) and $K(R)$,  
over the volume. 
The result is given in Fig.4 for the representative 
value $\delta=2.12$ (the results with $\delta$ 
between 1.9--2.2 being very similar); 
the model appears to be well
consistent with the data.

\section{Hard X--rays and EUV emission}

\subsection{The hard X--ray tail and EUV in the Coma cluster}

From a deep BeppoSAX observation, 
Fusco--Femiano et al.(1999) discovered 
an hard X--ray tail in the range 20--80 keV
exceeding the extrapolation 
of the thermal X-ray emission spectrum; 
this excess has also been confirmed by an RXTE 
observation (Rephaeli et al. 1999).
Due to the poor statistics, the spectrum is not well constrained
and the origin of this high 
energy emission cannot be firmly established.
It can be generated in non--thermal processes, possibly 
IC of the radio emitting electrons with the CMB photons
(Fusco--Femiano et al. 1999; Sarazin et al. 1999; Sarazin \&
Lieu 1998), or
it might originate from a supra--thermal power 
law tail of particles emitting via relativistic 
bremsstrahlung (Ensslin et al. 1999).
Furthermore, as recently shown in detail 
(Dogiel, these proceedings), 
it could also be of thermal origin (from a modification
of the pure maxwellian distribution of the hot cluster
gas induced by strong acceleration processes).

The IC origin is particularly fascinating in view of
the fact that a comparison between 
the radio emission from the halo and the hard X--ray tail
discovered by BeppoSAX allows to calculate the average
strength (over the cluster volume) 
of the magnetic field $B$ (Fusco--Femiano et al. 1999).

Since in the present model the break energy of the electron
population depends on the distance from the cluster center, 
the synchrotron as well as the IC spectra are emitted
by relativistic electrons whose
energy distribution cannot be represented by a unique power law 
over all the volume.
Therefore, the magnetic field cannot be simply evaluated by
standard formulae (e.g.Harris \& Grindlay 1979).
Under our assumptions 
the IC emissivity per unit energy and solid angle 
in the Thomson approximation is given by:

\begin{eqnarray}
j(\epsilon_1)=
K(R) {{r_0^2 \pi \epsilon_1}\over{8 c^2 h^3}} 
\int {{ d\epsilon}\over{ e^{\epsilon/ k T_{CMB}} -1}}
\int_{\gamma_{min}}^{\gamma_b(R)} 
{{d\gamma}\over{ \gamma^{\delta+4} }}
\cdot \nonumber\\
\left(1- {{\gamma}\over{\gamma_b(R)}} \right)^{\delta-2}
\left(
2\epsilon_1 ln {{\epsilon_1}\over{4\gamma^2 \epsilon}}
+\epsilon_1 +4\gamma^2 \epsilon 
-{{\epsilon_1^2}\over{2\gamma^2 \epsilon}} \right)
\end{eqnarray} 

where $\epsilon$ is the energy of the CMB photons and
for ultra--relativistic electrons 
$\gamma_{min} = \sqrt{\epsilon_1/ 4 \epsilon}$ 
(e.g. Blumenthal \& Gould 1970). The total
IC emission is readily obtained by integrating Eq.(7)
over the volume.

In Fig.5 we plot the expected IC flux from our model 
for different magnetic fields and compare the results with
the observations; the calculations are performed for
$\delta=2.12$ (in the model the value of $B$ is stable 
with $\delta \sim 1.9-2.2$).
\begin{figure}
 \resizebox{\hsize}{!}{\includegraphics{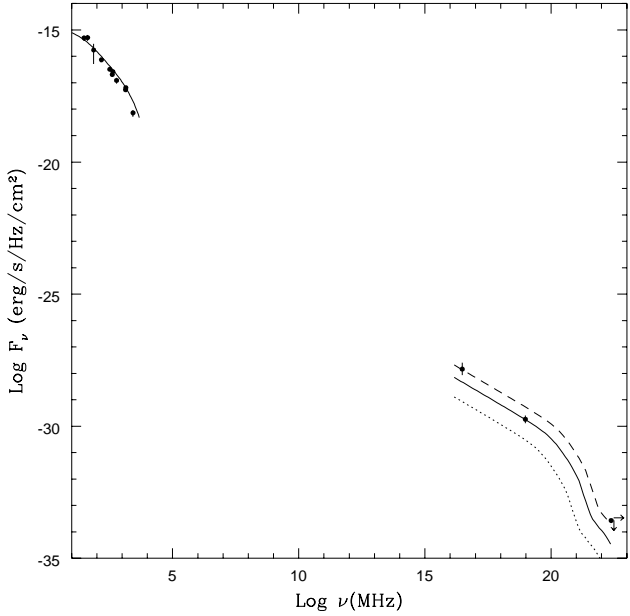}}
\caption[]{The expected radio and X--ray IC fluxes are
shown and compared with the radio  
(from Deiss et al.1997), EUV 
(Bowyer \& Bergh\"{o}fer 1998; Ensslin et al. 1999), 
hard X--ray (Fusco--Femiano 1999), and gamma--ray 
(Sreekumar et al. 1996) data.
The calculation of the IC flux ($\delta=2.12$)
is shown for different
magnetic field strengths: 0.5$\mu G$ (dotted line), 
0.1$\mu G$ (solid line), and
0.05$\mu G$ (dashed line).}
\end{figure}
Clearly the model cannot account
at the same time for the observed 
EUVs, X--rays and gamma--ray upper limits.
The X--ray 
data are well accounted for by a magnetic field
$B \sim 0.1 \mu G$, in agreement with 
the strength of a large scale magnetic field 
requested by Faraday rotation measurements (Feretti et al. 1995).
Furthermore, it should be noticed that 
the effect due to Coulomb losses on the 
electron energy distribution has not been taken into account in 
the IC calculation of Fig.5.
These losses depress the electron energy distribution at 
low energies ($\gamma < 300$) so that the estimated
IC emission in the EUV band should be considered as an upper limit 
(the predicted EUV flux 
would be $\sim$ 3 -- 5 times  
smaller than the observed value).

Due to the decrease of the synchrotron break 
frequency with $R$, in our model the observed 
radio FWHMs are expected to be
smaller than the predicted IC EUV FWHM. This is contrary to the
observational evidence
(Bowyer \& Berghoefer 1998; Ensslin et al. 1999), but
consistent with our IC model which underproduces by a large factor
the observed EUV flux.

\subsection{An additional electron population ?}

If the EUV excess is of non--thermal nature, 
an additional electron population is requested 
(Bowyer \& Berghoefer 1998; Ensslin et al. 1999).
It seems natural to assume an additional population of
relativistic electrons, 
with a relatively steep energy distribution ($\delta \simeq 2.8$), 
injected in the core of the Coma cluster 
by Active Galactic Nuclei (AGN).  

Since the physical conditions at each radius are the same for the two
populations ,
the break energies of this additional population
should be distributed as those 
of the synchrotron electrons (Fig.2, solid lines).
The normalization of the particle spectrum 
is obtained by imposing that it gives the observed EUV emission 
via IC scattering with the CMB photons. 
The radio spectral indices, radio brightness 
profiles, total synchrotron spectrum and hard X--ray IC emission
predicted by our model are not substantially modified by the
introduction of this additional electron population due to the
assumed steep spectral index.

\section{Conclusions}

We have studied the effect of a systematic reacceleration
on the relativistic plasma in the Coma halo.
Systematic reacceleration may
compensate for the radiative losses providing
a suitable mechanism to solve the problem
of the very short radiative life-times of the relativistic
electrons in the clusters.

We propose a two phase scenario in which the particles
are injected and reaccelerated in the cluster volume 
for a sufficiently long time ($\sim 1-3$ Gyrs), after 
which they are reaccelerated up to the present time.
The first phase (the injection phase) is probably related
to a major merger process in which relic particles 
(cosmic rays from galaxies and/or thermal particles) 
are (re)accelerated 
by shocks and/or induced strong turbulence.
During the second phase (post major merger phase) the 
electrons lose energy by emitting radiation and, at the same 
time, are reaccelerated by the cluster weather (i.e. 
galactic winds, accretion shocks, and/or turbulence).
The combination of losses and reacceleration 
produces flat synchrotron spectra in the central 
region and very steep spectra in the peripheral region 
well matching the observed radio spectral distribution in
the Coma halo. 
By assuming a constant value of $B$, 
the requested reacceleration times are relatively
large on the majority of the radio halo volume and the
reacceleration can be easily 
provided by systematic Fermi mechanisms.
Strong reaccelerations are requested only in the central 
region of the Coma halo 
(especially in the case of a small $B$ which is required
by the IC production of the observed 
hard X--ray tail). The problem of short acceleration times
can be eased by assuming a gradual increase of the magnetic
field strength 
toward the center of the cluster.
Since this would be necessary  
only in a small fraction of the cluster volume, 
our assumption of a constant magnetic field strength
over the cluster seems very reasonable. 

We show that the model can well reproduce the observed
brightness profiles at 327, 1380, and 1400 MHz, and the 
observed integrated synchrotron spectrum.
We have also shown that the model can account for the 
hard X--ray flux observed with BeppoSAX 
via the IC scattering of the CMB photons with
the synchrotron electrons
if a magnetic field
$B \sim 0.1 \mu G$ is assumed throughout the cluster medium.
In this case, the predicted hard X--ray spectral 
index is $\sim 0.6$ and 
the FWHM of the hard X--ray profile
is expected to be $\sim 30-35$ arcmin, i.e. 
considerably larger than the synchrotron FWHM at 
327 MHz; obviously, it would be very 
interesting to test these findings with future observations. 
The energy of the relativistic electrons reservoir
would be $\sim 5\cdot 10^{60}$erg s$^{-1}$, i.e.
$\sim 5\%$ of the thermal energy of the intracluster gas,
far greater than the energy in the magnetic field.

 The EUV excess from the 
Coma cluster cannot be accounted for by IC scattering 
of the synchrotron electrons with the CMB photons, the predicted
emissivity being a factor $3-5$ smaller
than observed.
If the EUV excess is non--thermal, then the presence
of an additional population of relativistic particles with
a more peaked radial distribution must be assumed.
We find that the 
synchrotron and IC hard X--ray contributions 
from an additional population of relativistic
electrons with a steep spectral index ($\delta=2.8$),
matching the EUV excess via IC scattering with the CMB photons,
do not substantially alter the fluxes estimated 
from the first population.
The additional population has a total energy in electrons 
similar to the first one and may have been injected by
AGNs activity within the cluster core during most
of the cluster life.

A much more detailed discussion about the time evolution 
of the relativistic plasma and the emitted synchrotron 
and IC spectra, including the
combined effects of Coulomb losses, systematic 
reacceleration and  
magnetic field dependence 
on $R$, will be presented in a forthcoming paper
(Brunetti et al. 1999).

\begin{acknowledgements}
We are greatful to A.Atoyan, S.Colafrancesco, R.Ekers, T.Ensslin  
 for useful discussions during 
the Ringberg workshop.
This work was partly supported by the Italian Ministry for
University and Research (MURST) under grant Cofin98-02-32, 
and by the Italian Space Agency (ASI).
\end{acknowledgements}

\end{document}